\documentclass[draft,journal,transmag,onecolumn]{IEEEtran}
\usepackage{ifpdf}

\usepackage{cite}

\ifCLASSINFOpdf
\else
\fi
\usepackage{amsmath}
\usepackage{amssymb}
\usepackage{bm}
\usepackage{bbm}
\usepackage{mathtools}
\usepackage{bigints}

\newtheorem{thm}{Theorem}[section]
\newtheorem{lem}[thm]{Lemma}

\newtheorem{ex}{Example}

\begin{document}
%
\title{A Revisitation of Low-Rate Bounds on the Reliability Function of Discrete Memoryless Channels for List Decoding}


\author{\IEEEauthorblockN{Marco Bondaschi and
Marco Dalai,~\IEEEmembership{Senior Member,~IEEE}}
\thanks{M. Bondaschi is with the School of Computer and Communication Sciences, École Polytechnique Fédérale de Lausanne, CH-1015 Lausanne, Switzerland (e-mail: marco.bondaschi@epfl.ch).
M. Dalai is with the Department of Information Engineering at the University of Brescia, Via
Branze 38, I-25123 Brescia, Italy (e-mail:  marco.dalai@unibs.it).
Part of this work was presented at ISIT 2020. }}

%



\IEEEtitleabstractindextext{%
\begin{abstract}
We revise the proof of low-rate upper bounds on the reliability function of discrete memoryless channels for ordinary and list-decoding schemes, in particular Berlekamp and Blinovsky's zero-rate bound, as well as Blahut's bound for low rates. The available proofs of the zero-rate bound devised by Berlekamp and Blinovsky are somehow complicated in that they contain in one form or another some cumbersome ``non-standard'' procedures or computations. Here we follow Blinovsky's idea of using a Ramsey-theoretic result by Koml\'{o}s, and we complement it with some missing steps to present a proof which is rigorous and easier to inspect. Furthermore, we show how these techniques can be used to fix an error that invalidated the proof of Blahut's low-rate bound, which is here presented in an extended form for list decoding and for general channels.
\end{abstract}

}

\maketitle

\IEEEdisplaynontitleabstractindextext

%
\IEEEpeerreviewmaketitle

\section{Introduction}
%
%
%
%
We consider a discrete memoryless channel with input alphabet $\mathcal{X}=\{1,2, \dots, \lvert\mathcal{X}\rvert\}$, discrete output alphabet $\mathcal{Y}$ and transition probabilities $P(y|x)$. An $L$-list coding scheme with message set $\mathcal{M}=\{1,2,\ldots,M\}$ and blocklength $n$ is composed of an encoder $\mathcal{C} : \mathcal{M}  \to \mathcal{X}^n$ and a  decoder $\mathcal{C}^{-1} : \mathcal{Y}^n \to [\mathcal{M}]^L
$, where the symbol $[\mathcal{M}]^L$ denotes the set of all subsets of $\mathcal{M}$ of cardinality $L$. The rate of transmission is defined as $R=\log(M/L)/n$.

In this paper, $L$ is to be considered as a fixed parameter. The setting we are interested in is the classical one where $M$ grows exponentially in $n$ according to a fixed rate, that is we will consider a fixed $R$ and let $M$ be the least integer greater than or equal to\footnote{We follow \cite{sgb1} in the definition of the rate. Other works, such as \cite{blinovsky1} for example, define $R=\log(M)/n$. As it will later become clear, this has no impact on the resulting bounds for a fixed $L$.} $ L\exp(nR)$. 

When message $m$ is sent, an output sequence $\mathbf{y}=(y_1,y_2,\ldots, y_n)$ is received with probability 
\begin{equation}
P_m(\mathbf{y}) = \prod_{i=1}^n P(y_i|x_{m,i})\,,
\end{equation}
where $\mathbf{x}_m=(x_{m,1},x_{m,2},\ldots,x_{m,n})=\mathcal{C}(m)$.
The decoder, after receiving $\mathbf{y}$, produces a list $\mathcal{C}^{-1}(\mathbf{y})$ of $L$ messages and an error occurs if $m\not \in \mathcal{C}^{-1}(\mathbf{y})$.
This happens with probability 
\begin{equation}
P_{e,m} = \sum_{\mathbf{y} \in \mathsf{Y}_m^c} P_m(\mathbf{y})\,,
\end{equation}
where $\mathsf{Y}_m \subset \mathcal{Y}^n$ is the subset of output sequences whose decoded list $\mathcal{C}^{-1}(\mathbf{y})$ contains $m$. The average probability of error of the code, when the messages are sent with equal probability, is 
\begin{equation}
P_{e} = \frac{1}{M} \sum_{m=1}^M P_{e,m}\,.
\end{equation}
The decoding scheme achieving the smallest probability of error is the maximum-likelihood decoder, which for any given sequence $\mathbf{y}$ outputs a list containing the $L$ messages $m$ with the largest $P_m(\mathbf{y})$. Ties can be resolved arbitrarily, since they do not affect the overall probability of error.
It can be seen that $P_{e}>0$ if any set of $L+1$ channel inputs may produce the same output with non-zero probability, that is, the zero-error capacity with list-size $L$ is zero \cite{elias1}. We will assume this through the whole paper.

For fixed $R$, $n$ and $L$, let $P_e(L,R,n)$ be the smallest probability of error for $L$-list decoding over all codes with rate at least $R$ and block length $n$. The reliability function is defined as 
\begin{equation}
\label{ELDef}
E_L(R) \triangleq \limsup_{n \to \infty} -\frac{\log P_e(L,R,n)}{n}\,.
\end{equation}

It is known \cite{sgb1, gallager2} that the same function is obtained if one replaces $P_e$ with the \emph{maximal} probability of error $
P_{e,\max} = \max_{m} P_{e,m}$. In this paper we will lower bound $P_{e,\max}$ to derive upper bounds on $E_L(R)$. In particular, the main focus is on bounding the limiting value as $R$ approaches $0$, say $E_L(0^+)$. This will later also be useful for bounds on $E_L(R)$ at $R>0$.

It is known that $E_L(0^+)$ has the single-letter expression 
\begin{equation}
\label{EL(0)_value}
E_L(0^+) = \max_{Q \in \mathcal{P}(\mathcal{X})}\bigg[-\sum_{\mathbf{x} \in \mathcal{X}^{L+1}} Q(x_1)\cdots Q(x_{L+1}) \log \sum_{y \in \mathcal{Y}} \sqrt[L+1]{P(y|x_1)\cdots P(y|x_{L+1})}\bigg]\,,
\end{equation}
where $\mathcal{P}(\mathcal{X})$ is the set of all probability distributions on $\mathcal{X}$.
For $L=1$, the achievability part was proved by Gallager \cite{gallager2} using his expurgated bound while the converse was proved by Berlekamp in his doctoral thesis \cite{berlekamp1} and published (with some misprints) in \cite{sgb2}. The achievability is easily extendable to $L>1$ while the converse was extended by Blinovsky in \cite{blinovsky1}. While the achievability part is rather well understood and different equivalent ways of deriving the results are known \cite{somekh1, merhav1, omura1} (perhaps explicitly mentioned usually for $L=1$), the converse remains more obscure. The original proof in \cite{sgb2} is based on a rather unusual procedure which involves recursive decomposition and concatenation of codes (see also \cite[Problems 10.20-10.21]{csiszar1} for a summary of the proof, as well as \cite{wagner1} for an alternative formulation of the procedure in terms of inner-product spaces).
A simpler procedure was proposed in \cite{blahut} (see also \cite{blahut2}), but unfortunately the proof contains a gap (see \cite[Sec. VI.A]{dalai-2015} for details) which limits the application to pairwise symmetric channels, for which a simple proof was already mentioned in \cite{sgb2}. For fixed $L>1$, the proof used by Blinovsky in \cite{blinovsky1} follows the same idea employed in Berlekamp's original proof. A simpler proof was later sketched in \cite{blinovsky2} (for $L=1$) and in \cite{ahlswede1} (for $L>1$). The main key ingredient in those simplifications is the use of a result by Koml\'os \cite{komlos1} in Ramsey theory, which proved to be an important tool for results on low rate codes (see for example \cite{polyanskiy, alon, wang, zhang} for recent results and discussions on the use of these methods in a similar context of list decoding for adversarial channels).

The proofs in \cite{blinovsky1, ahlswede1} still share some steps that require some troublesome multivariate analysis, and because of this they turn out to be difficult to inspect. In this paper, we fill in all missing steps to extend Blinovsky's simplification in an effective way to prove the converse part of \eqref{EL(0)_value} for general $L>1$. We do not claim originality of the used ideas; quite to the contrary, we believe our contribution is precisely to sort out ideas scattered in different works and complement them with some standard ones to produce a rigorous and flexible proof for the case of general $L$. As an implication of this revisitation, we show in the last part of the paper that Blahut's proof of his upper bound on $E_L(R)$ for $R\geq0$ and $L=1$ can be fixed easily using the presented setting and also extended to the case of general $L\geq 1$. Another example of application of this way of looking at the problem can be found in \cite{bondaschi}, where the case of mismatched decoding is considered.

The rest of the paper is organized as follows. In Section II we study the probability of error for $M=L+1$ codewords using the method of types, and reduce the problem for general $M$ to that for the worst subset of $L+1$ codewords. In Section III we employ Koml\'os' results \cite{komlos1} to extract a subcode that satisfies certain symmetric properties. Using this symmetry, we prove the upper bound on $E_L(0^+)$, in  Section IV, employing the usual Plotkin-like double counting trick. Finally, in Section \ref{sec:Blahut} we discuss the upper bound on the error exponent proposed by Blahut in \cite{blahut}.

\section{Probability of Error and Minimum Discrepancy}

We derive in this section a bound on the error probability for a given code based on a measure of discrepancy between codewords, taken in groups of $L+1$ of them at a time. The obtained bound will depend here on the specific structure of the code. Later, we will show how to extract from codes of positive rate subcodes with a symmetry which allows one to derive an effective bound first on the discrepancy, and consequently on the probability of error, which does not depend on the code.

Consider first a fixed set of $M=L+1$ codewords $\{\mathbf{x}_1,\dots,\mathbf{x}_{L+1}\}$, where $L$ is the list size. As we already pointed out in the introduction, the decoding scheme achieving the smallest probability of error is the maximum-likelihood decoder, for which $\mathsf{Y}_m^c$ contains sequences $\mathbf{y}$ such that $\min_i P_i(\mathbf{y}) = P_m(\mathbf{y})$. This is due to the fact that, since $M=L+1$, only one message is left out of the list for each $\mathbf{y}$. If the messages with the smallest probability are more than one, then any of them can be left out of the list without affecting the overall probability of error. We want to group together sequences $\mathbf{y}$ that have the same $P_m(\mathbf{y})$ for any $m$, since they can be decoded in the same way without affecting $P_e$. To this end, consider a generalization of conditional types as defined by Csisz\'{a}r and K\"{o}rner \cite{csiszar1}, in which instead of a single codeword conditioning the output sequences $\mathbf{y}$, we consider the whole set of $L+1$ codewords. Therefore, in this setting an ``input symbol'' is any of the possible $\lvert \mathcal{X}\rvert^{L+1}$ sequences of $L+1$ input symbols from $\mathcal{X}$, i.e., any element of $\mathcal{X}^{L+1}$. If we imagine the code $\mathcal{C}$ as an $(L+1)\times n$ matrix, to each sequence $\mathbf{x} \in \mathcal{X}^{L+1}$ we can associate a region of coordinates $I_{\mathbf{x}}$, which is the set of coordinates where the code has the sequence $\mathbf{x}$ as a column. Then to each output sequence $\mathbf{y}$ we can associate its conditional type $T(\mathbf{y}): \mathcal{X}^{L+1} \to \mathcal{P}(\mathcal{Y})$ that assigns to each $\mathbf{x}\in\mathcal{X}^{L+1}$ a probability distribution $T_{\mathbf{x}}(\mathbf{y})$ on $\mathcal{Y}$ such that each $y\in\mathcal{Y}$ has a probability equal to the fraction of times the symbol $y \in \mathcal{Y}$ occurs in $\mathbf{y}$ in the region of coordinates $I_{\mathbf{x}}$ (that is, $T_{\mathbf{x}}(\mathbf{y})$ is the empirical distribution of $\mathbf{y}$ restricted to the set of coordinates $I_{\mathbf{x}}$). Notice that all output sequences $\mathbf{y}$ that have the same conditional type given the whole code, also have the same $P_m(\mathbf{y})$ for any message $m$.

\begin{ex}
Suppose we have binary alphabets $\mathcal{X}=\mathcal{Y}=\{0,1\}$ and a code with $M=3$ codewords of block length $n=10$:
\begin{align*}
\mathbf{x}_1 &= 0000011111 \\
\mathbf{x}_2 &= 0101010101 \\
\mathbf{x}_3 &= 0101011111.
\end{align*}
For these codewords, we only have 4 sequences of input symbols that appear at one or more coordinates: $(0,0,0)$ at coordinates $1$, $3$ and $5$, $(0,1,1)$ at coordinates $2$ and $4$, $(1,0,1)$ at coordinates $7$ and $9$, and $(1,1,1)$ at coordinates $6$, $8$ and $10$. The corresponding sets of coordinates are $I_{000}$, $I_{011}$, $I_{101}$ and $I_{111}$; to all the other sequences of input symbols correspond empty sets of coordinates. Furthermore, we have $n_1=n_4=3$ and $n_2=n_3=2$.

Consider now the output sequence
\begin{equation*}
\mathbf{y} = 0101110011\,.
\end{equation*}
In the first set of coordinates $I_{000}$, i.e., coordinates $1$, $3$ and $5$, symbol $0$ occurs 2 out of 3 times, and symbol $1$ occurs 1 out of 3 times; hence, $T_{000}(\mathbf{y})=\big(\frac{2}{3},\frac{1}{3}\big)$. The same reasoning applies to the other three sets of coordinates, yielding $T_{011}(\mathbf{y}) = (0,1)$, $T_{101}(\mathbf{y}) = \big(\frac{1}{2},\frac{1}{2}\big)$ and $T_{111}(\mathbf{y}) =\big(\frac{1}{3},\frac{2}{3}\big)$. Any other output sequence with the same type as $\mathbf{y}$, for example
\begin{equation*}
\mathbf{y}^* = 1101011100\,,
\end{equation*}
has the same probabilities $P_1$, $P_2$ and $P_3$ as $\mathbf{y}$, regardless of the discrete memoryless channel under consideration. \null\hfill $\square$
\end{ex}

In the following, we will use the basic fact about conditional types \cite[Lemma 2.6]{csiszar1} that for any given probability distribution $Q$ on $\mathcal{Y}^n$ in the form
\begin{equation}
\label{PDConstraint}
Q(\mathbf{y}) = \prod_{\mathbf{x} \in \mathcal{X}^{L+1}} \prod_{i \in I_{\mathbf{x}}} Q_{\mathbf{x}}(y_i)\,,
\end{equation}
where for all $\mathbf{x}$, $Q_{\mathbf{x}}(y)$ is a probability distribution on $\mathcal{Y}$, we have that
\begin{equation}
\label{QLower}
Q(T) \triangleq \sum_{\mathbf{y} \,:\, T(\mathbf{y}) = T} Q(\mathbf{y}) \geq \frac{1}{(n+1)^{\lvert\mathcal{X}\rvert^{L+1}\lvert\mathcal{Y}\rvert}} \exp\Bigg\{-n\sum_{\mathbf{x} \in \mathcal{X}^{L+1}} q(\mathbf{x}) D(T_{\mathbf{x}}(T) \,||\, Q_{\mathbf{x}})\Bigg\}\,,
\end{equation}
where we denoted (with a slight abuse of notation) by $Q(T)$ the probability (under $Q$) of the (non-empty) set of all sequences $\mathbf{y}$ of conditional type $T$, by $T_{\mathbf{x}}(T)$ the probability distribution $T_{\mathbf{x}}(\mathbf{y})$ of any output sequence $\mathbf{y}$ of type $T(\mathbf{y})=T$, and by $q(\mathbf{x})$ the fraction of times the sequence $\mathbf{x}$ occurs as a column in the whole code (that is, the joint type of the $L+1$ codewords) and where $D(P\,||\,Q)$ is the Kullback-Leibler divergence between distributions $P$ and $Q$ defined on the same discrete alphabet $\mathcal{A}$,
\begin{equation}
D(P\,||\,Q) \triangleq \sum_{a\in\mathcal{A}} P(a)\log\frac{P(a)}{Q(a)}\,.
\end{equation}

We can now proceed to study the overall probability of error for the maximum-likelihood $L$-list decoding scheme, when $M=L+1$. For all $m \in \mathcal{M}$ we have
\begin{equation}
\mathbf{y} \in \mathsf{Y}_m^c \implies \log\frac{P_k(\mathbf{y})}{P_m(\mathbf{y})} \geq 0 \quad\forall\, k \in \mathcal{M}.
\end{equation}
The last implication can be rewritten with some manipulations as: $\mathbf{y} \in \mathsf{Y}_m^c$ implies that
\begin{equation*}
\sum_{\mathbf{x}} q(\mathbf{x}) D(T_{\mathbf{x}}(\mathbf{y})\, ||\, P(\cdot | x_m)) \geq \sum_{\mathbf{x}} q(\mathbf{x}) D(T_{\mathbf{x}}(\mathbf{y})\,||\, P(\cdot | x_k))
\end{equation*}
for all $k \in \mathcal{M}$, where $x_k$ is the $k$-th symbol of $\mathbf{x}$, and $P(\cdot |x_k)$ is the probability distribution on the outputs of the channel given the input symbol $x_k$.
Hence, we can see the decoding regions as decoding regions on types instead of sequences, and rewrite the implication above as: $T \in \mathcal{T}_m^c$ implies that
\begin{equation}
\label{TypesDecoding}
\max_{k \in \mathcal{M}}\, \sum_{\mathbf{x}} q(\mathbf{x}) D(T_{\mathbf{x}}(T) \,||\, P(\cdot | x_k)) = \sum_{\mathbf{x}} q(\mathbf{x}) D(T_{\mathbf{x}}(T) \,||\, P(\cdot | x_m))\,,
\end{equation}
where $\mathcal{T}_m$ is the set of types of the sequences decoded to a list that includes message $m$.
In order to avoid that some of the KL divergences go to infinity, in the following we will consider only the output sequences $\mathbf{y}$ belonging to the set
\begin{equation}
\label{YnHat}
\hat{\mathcal{Y}}^n = \{\mathbf{y} \in \mathcal{Y}^n \mid P_1(\mathbf{y})P_2(\mathbf{y})\cdots P_{L+1}(\mathbf{y}) > 0\}\,.
\end{equation}
Note that this set is non-empty for any code since we are assuming that the channel under consideration has zero-error capacity (for $L$-list decoding) $C_{0,L} = 0$. Also, there is no loss of generality in limiting the attention to $\hat{\mathcal{Y}}^n$, since if $P_m(\mathbf{y})=0$ for some $m$, then sequence $\mathbf{y}$ is always ($L$-list) decoded correctly and it does not contribute to $P_e$. This also means that for all the coordinates belonging to region $I_{\mathbf{x}}$ we consider only the output symbols $y$ that belong to the set 
\begin{equation}
\label{YxHat}
\hat{\mathcal{Y}}_{\mathbf{x}} = \{y \in \mathcal{Y} \mid P(y|x_1)P(y|x_2) \dots P(y|x_{L+1}) > 0\}\,.
\end{equation}
So, from now on we consider that types $T_{\mathbf{x}}$ are constrained to have components equal to $0$ at all $y\notin \hat{\mathcal{Y}}_{\mathbf{x}}$.

Then, the average probability of error of the $L+1$ codewords of length $n$ is:
\begin{equation}
P_e = \frac{1}{L+1} \sum_{m=1}^{L+1} P_{e,m} = \frac{1}{L+1} \sum_{m=1}^{L+1} \sum_{T \in \mathcal{T}_m^c} P_m(T)\,, \label{PeInt}
\end{equation}
where
\begin{equation}
P_m(T) \triangleq \sum_{\mathbf{y} \,:\, T(\mathbf{y}) = T} P_m(\mathbf{y})\,.
\end{equation}
Since all $P_m$ are in the form \eqref{PDConstraint}, we can use the lower bound \eqref{QLower} and \eqref{TypesDecoding} to get:
\begin{align}
P_e & \geq \sum_{m=1}^{L+1} \sum_{T \in \mathcal{T}_m^c} \exp\Bigg\{\!-n \bigg(\sum_{\mathbf{x}} q(\mathbf{x}) D(T_{\mathbf{x}}(T) \,||\, P(\cdot | x_m)) +o(1)\bigg)\Bigg\} \notag \\
	& = \sum_{T \in \mathcal{T}(\mathcal{C})} \exp\Bigg\{\!-n \bigg(\max_{k \in \mathcal{M}} \sum_{\mathbf{x}} q(\mathbf{x}) D(T_{\mathbf{x}}(T) \,||\, P(\cdot |x_k))+o(1)\bigg)\Bigg\} \notag \\
	& \geq \exp \Bigg\{\!-n\bigg(\min_{T \in \mathcal{T}(\mathcal{C})}\max_{k \in \mathcal{M}} \sum_{\mathbf{x}} q(\mathbf{x}) D(T_{\mathbf{x}}(T)\,||\, P(\cdot |x_k))+o(1)\bigg)\Bigg\}\,, \label{PeLower}
\end{align}
where $T(\mathcal{C})$ is the set of all possible conditional types given the code $\mathcal{C}$, and $o(1)$ is a quantity which vanishes as $n\to\infty$, which only depends on $L$, $|\mathcal{X}|$ and $|\mathcal{Y}|$ but not on the codewords.

We now analyze more closely the sum at the exponent in equation \eqref{PeLower}. We first observe that we can replace types with general distributions with some additional $o(1)$ penalty. More specifically, we replace the minimization over $\mathcal{T}(\mathcal{C})$ with one over the cartesian product $\prod_{\mathbf{x}\in\mathcal{X}^{L+1}}\mathcal{P}(\hat{\mathcal{Y}}_{\mathbf{x}})$, where
 $\mathcal{P}(\hat{\mathcal{Y}}_{\mathbf{x}})$ is the set of all probability distributions on $\hat{\mathcal{Y}}_{\mathbf{x}}$. Indeed, for any $\mathbf{x}$, since $T_{\mathbf{x}}(y)=0$ whenever $P(y|x_k)=0$, we have
\begin{align*}
 D(T_{\mathbf{x}}(T)\,||\, P(\cdot |x_k)) & = \sum_y T_{\mathbf{x}}(y)\log\frac{1}{P(y|x_k)}-H(T_{\mathbf{x}}(T))\\
& \leq \log\frac{1}{P_{\min}}\,,
\end{align*}
where $P_{\min}$ is the smallest non-zero transition probability. So, if say $q(\mathbf{x})\leq 1/\sqrt{n}$ then the contribution of $q(\mathbf{x}) D(T_{\mathbf{x}}(T)\,||\, P(\cdot |x_k))$ in the sum is $o(1)$. On the other hand, if $q(\mathbf{x})>1/\sqrt{n}$ then any distribution $T\in  \mathcal{P}(\hat{\mathcal{Y}}_{\mathbf{x}})$ is approximated with error at most $1/\sqrt{n}$ in any component by some type $T_{\mathbf{x}}$ (see \cite[pag. 18]{shannon1}). So, by continuity of the function $t\log t$, the contribution of such $\mathbf{x}$ to the minimum over types will differ from that given to the minimum over distributions by a $o(1)$ difference. Hence, we reach the conclusion that
$$
P_e\geq e^{-n (D_{\mathcal{M}} +o(1))}\,,
$$
where
\begin{equation}
D_{\mathcal{M}} \triangleq \min_{T \in \prod\limits_{\mathbf{x}}\mathcal{P}(\hat{\mathcal{Y}}_{\mathbf{x}})} \max_k \sum_{\mathbf{x}} q(\mathbf{x}) D(T_{\mathbf{x}}(T)\,||\, P(\cdot |x_k))
\end{equation}
and $o(1)$ now also depends on the channel (but not on the coderwords).
If we introduce a vector $\bm{\alpha} = (\alpha_1,\alpha_2,\dots,\alpha_{L+1})$ with $\alpha_k \geq 0$ and $\sum_k \alpha_k = 1$, we can also write that
\begin{equation}
\label{Exponent}
D_{\mathcal{M}} = \min_{T \in \prod\limits_{\mathbf{x}}\mathcal{P}(\hat{\mathcal{Y}}_{\mathbf{x}})} \max_{\bm{\alpha}} \sum_{k=1}^{L+1} \alpha_k \sum_{\mathbf{x}} q(\mathbf{x}) D(T_{\mathbf{x}}(T)\,||\, P(\cdot |x_k))
\end{equation}
since the maximum over $\bm{\alpha}$ is obtained when the weight is all on the largest KL divergence. Since the set over which we take the minimum is convex and compact, the set $\{\bm{\alpha}\}$ is convex, and the objective function in \eqref{Exponent} is linear in $\bm{\alpha}$ for any $T$ and it is convex and lower semi-continuous in $T$ for any $\bm{\alpha}$, by \cite[Theorem 4.2']{sion1} the $\min$ and $\max$ can be exchanged, leading to
\begin{equation}
\label{MinMaxExchange}
D_{\mathcal{M}} = \max_{\bm{\alpha}} \min_{T \in \prod\limits_{\mathbf{x}}\mathcal{P}(\hat{\mathcal{Y}}_{\mathbf{x}})}\sum_{k=1}^{L+1} \alpha_k \sum_{\mathbf{x}} q(\mathbf{x}) D(T_{\mathbf{x}}(T)\,||\, P(\cdot |x_k))\,.
\end{equation}
We can now apply the following lemma.
\begin{lem}[Shayevitz \cite{shayevitz1}] 
Let $P_1,P_2,\dots,P_{K}$ be $K$ probability distributions on a finite alphabet $\mathcal{A}$. Let also
\begin{equation}
\mu(\bm{\alpha}) \triangleq -\log \sum_{a \in\mathcal{A}} P_1(a)^{\alpha_1} \cdots P_{K}(a)^{\alpha_{K}}\,,
\end{equation}
where $\alpha_i \geq 0$ and $\sum_i \alpha_i = 1$. Then,
\begin{equation}
\mu(\bm{\alpha}) = \min_{Q \in \mathcal{P}(\mathcal{A})} \sum_{k=1}^{K} \alpha_k D(Q \,||\,P_k)\,.
\end{equation}
\null\hfill $\square$
\end{lem}

In our case, let
\begin{equation}
\label{MuFirstDef}
\mu_{\mathcal{M}}(\bm{\alpha}) \triangleq -\log \sum_{\mathbf{y} \in \hat{\mathcal{Y}}^n} P_1(\mathbf{y})^{\alpha_1} \cdots P_{L+1}(\mathbf{y})^{\alpha_{L+1}}.
\end{equation}
From Lemma 2.1 it follows that
\begin{equation}
\label{MuMinDef}
\mu_{\mathcal{M}}(\bm{\alpha}) = \min_{Q \in \mathcal{P}(\mathcal{Y}^n)} \sum_{k=1}^{L+1} \alpha_k D(Q \,||\,P_k)\,.
\end{equation}
It can be verified by substitution that the distribution $Q$ that minimizes this expression is
\begin{equation}
\label{QStar}
Q^*(\mathbf{y}) = \frac{P_1(\mathbf{y})^{\alpha_1} \cdots P_{L+1}(\mathbf{y})^{\alpha_{L+1}}}{\sum_{\mathbf{y}' \in \hat{\mathcal{Y}}^n} P_1(\mathbf{y}')^{\alpha_1} \cdots P_{L+1}(\mathbf{y}')^{\alpha_{L+1}}}\,.
\end{equation}

Now, $Q^*(\mathbf{y})$ can be put in the form \eqref{PDConstraint}, due to the fact that both numerator and denominator can be factorized symbol-wise, i.e., $Q^*(\mathbf{y}) = \prod_{\mathbf{x}}\prod_{i\in I_{\mathbf{x}}}Q^*_{\mathbf{x}}(y_i)$, with
\begin{equation}
Q^*_{\mathbf{x}}(y) \triangleq \frac{P(y|x_1)^{\alpha_1} \cdots P(y|x_{L+1})^{\alpha_{L+1}}}{\sum_{y' \in \hat{\mathcal{Y}}_{\mathbf{x}}} P(y'|x_1)^{\alpha_1} \cdots P(y'|x_{L+1})^{\alpha_{L+1}}}
\end{equation}
and for all $\mathbf{x}$, $Q^*_{\mathbf{x}}(y)$ also belongs to $\mathcal{P}(\hat{\mathcal{Y}}_{\mathbf{x}})$. Furthermore, for all probability distributions in the form \eqref{PDConstraint} we have, due to the additivity of the KL divergence for product distributions,
\begin{equation}
\label{KLAdditivity}
D(Q\,||\, P_k) = n \sum_{\mathbf{x}} q(\mathbf{x}) D(Q_{\mathbf{x}} \,||\, P(\cdot | x_k))\,,
\end{equation}
where $Q_{\mathbf{x}}$ is the probability distribution on $\mathcal{Y}$ that comes from the factorization of $Q$ according to \eqref{PDConstraint}.
Hence, from \eqref{MuMinDef} and \eqref{KLAdditivity} it follows that
\begin{equation}
\label{MuMinDef2}
\frac{1}{n}\,\mu_{\mathcal{M}}(\bm{\alpha}) = \min_{T \in \prod\limits_{\mathbf{x}}\mathcal{P}(\hat{\mathcal{Y}}_{\mathbf{x}})}\sum_{k=1}^{L+1} \alpha_k \sum_{\mathbf{x}} q(\mathbf{x}) D(T_{\mathbf{x}}(T)\,||\, P(\cdot |x_k))
\end{equation}
and therefore we have from \eqref{MinMaxExchange} and \eqref{MuMinDef2} that
\begin{equation}
\label{DMDef}
D_{\mathcal{M}} = \frac{1}{n} \max_{\bm{\alpha}} \mu_{\mathcal{M}}(\bm{\alpha})\,.
\end{equation}
Next, if we define, for the sequence of input symbols $\mathbf{x}=(x_1,x_2,\dots,x_{L+1})$, the function
\begin{equation}
\label{MuX}
\mu_{\mathbf{x}}(\bm{\alpha}) \triangleq -\log \sum_{y \in \mathcal{Y}} P(y|x_1)^{\alpha_1} \cdots P(y|x_{L+1})^{\alpha_{L+1}}\,,
\end{equation}
then we can use the additivity of $\mu_{\mathcal{M}}(\bm{\alpha})$,
\begin{equation}
\mu_{\mathcal{M}}(\bm{\alpha}) = \sum_{i=1}^n \mu_i(\bm{\alpha})\,,
\end{equation} 
where
\begin{equation}
\mu_i(\bm{\alpha}) \triangleq -\log \sum_{y \in \hat{\mathcal{Y}}_i} P(y|x_{1,i})^{\alpha_1} P(y|x_{2,i})^{\alpha_2} \cdots P(y|x_{L+1,i})^{\alpha_{L+1}}
\end{equation}
which, again, follows from the fact that the sum in \eqref{MuFirstDef} can be factorized symbol-wise, to rewrite \eqref{MuFirstDef} as
\begin{equation}
\label{MuLetters}
\mu_{\mathcal{M}}(\bm{\alpha}) = n \sum_{\mathbf{x} \in \mathcal{X}^{L+1}} q(\mathbf{x}) \,\mu_{\mathbf{x}}(\bm{\alpha})
\end{equation}
by grouping together the (equal) elements of the sum corresponding to the same region $I_{\mathbf{x}}$.
So, from the discussion above we conclude that for any $\delta > 0$ there exists a $n_0$, which only depends on the channel and on $L$, such that for any code of length $n\geq n_0$, wth $L+1$ codewords,
\begin{equation}
\label{PeBound}
P_{e,\max} \geq P_e \geq \exp \big\{-n (D_{\mathcal{M}} + \delta)\big\}\,,
\end{equation}
where
\begin{equation}
D_{\mathcal{M}}= \max_{\bm{\alpha}}\sum_{\mathbf{x} \in \mathcal{X}^{L+1}} q(\mathbf{x}) \,\mu_{\mathbf{x}}(\bm{\alpha})\,.
\end{equation}
The quantity $D_{\mathcal{M}}$ can be interpreted as a measure of \emph{discrepancy} for a set $\mathcal{M}$ of $L+1$ codewords.

If we now consider a fixed code $\mathcal{C}$ with $M \geq L+1$ messages $\mathcal{M}=\{1,2,\dots,M\}$, for any subset of $L+1$ messages $\mathbf{m} \in \mathcal{M}$ we have, by equation \eqref{PeBound}, that for any $\delta>0$ there exists a $n_0$ such that for at least one message $m \in \mathbf{m}$,
\begin{equation}
P_{e,m} \geq \exp \big\{-n (D_{\mathbf{m}} + \delta)\big\} \qquad \forall\,n \geq n_0,
\end{equation}
where 
\begin{equation}
D_{\mathbf{m}}\triangleq \max_{\bm{\alpha}}\sum_{\mathbf{x} \in \mathcal{X}^{L+1}} q_{\mathbf{m}}(\mathbf{x}) \,\mu_{\mathbf{x}}(\bm{\alpha})
\end{equation}
and $q_{\mathbf{m}}(\mathbf{x})$ is the fraction of times the sequence $\mathbf{x}$ occurs in the same coordinate in the codewords of the messages $\mathbf{m}$  (that is, the joint type of the $L+1$ codewords associated to $\mathbf{m}$).
If we define the \emph{minimum discrepancy} of the code $\mathcal{C}$ as
\begin{equation}
\label{Dmin}
D_{\min}(\mathcal{C}) \triangleq \min_{\mathbf{m} \subset \mathcal{M}} D_{\mathbf{m}}\,,
\end{equation}
where the minimum is over all $(L+1)$-subsets of $\mathcal{M}$, then for some message $\hat{m}$ we have 
\begin{equation}
P_{e,\hat{m}} \geq \exp \big\{-n (D_{\min}(\mathcal{C}) + \delta)\big\} \qquad \forall\,n \geq n_0
\end{equation}
and therefore, the maximal probability of error of $\mathcal{C}$ is lower bounded by
\begin{equation}
\label{PeMLower}
P_{e,\max} \geq P_{e,\hat{m}} \geq \exp \big\{-n \big(D_{\min}(\mathcal{C}) +\delta\big)\big\} \qquad \forall\,n \geq n_0,
\end{equation}
where again $n_0$ only depends on the channel and on $L$, and not on the code $\mathcal{C}$.

The way we derived the lower bound above for general $L$ is different from those in \cite{sgb2, blinovsky1, ahlswede1}; the advantage of our approach is that it does not require the study of the behaviour of the gradient of $\mu_{\mathcal{M}}(\bm{\alpha})$ on the border of the set $\{\bm{\alpha}\}$, which turns out to be rather tedious for $L>1$.

\section{Ramsey Theory and Generalization of Koml\'{o}s' result}
\label{KomlosSection}

Consider a set of $M$ random variables $\{X_1,X_2, \dots, X_M\}$ with indices in $\mathcal{M} = \{1,2,\dots,M\}$, taking values in a finite alphabet $\mathcal{X}$. Let us call $\chi_m(x) \triangleq \chi(X_m = x)$ the indicator function over the sample space $\Omega$ of the random variable $X_m$ taking the value $x \in \mathcal{X}$. With this notation,
\begin{gather*}
\int \chi_m(x) \,dP = P(X_m = x)\,, \\
\int \chi_m(x) \chi_{m'}(x') \,dP = P(X_m=x, X_{m'}=x')\,,
\end{gather*}
and so on.
Define also the averages
\begin{equation}
\label{AverageDef}
\overline{\chi}_m(x) \triangleq \frac{\chi_1(x) + \chi_2(x) + \dots + \chi_m(x)}{m}\,.
\end{equation}
Then, the following lemma holds.

\begin{lem}[Koml\'{o}s \cite{komlos1}]
\label{KomlosLemma1}
If for a fixed $x \in \mathcal{X}$ there exists a number $r_x$ such that for all $m<m' \in \mathcal{M}$,
\begin{equation}
\label{KomlosLemma1Cond}
\bigg\lvert \int \chi_m(x) \chi_{m'}(x) \,dP - r_x\,\bigg\rvert \leq \varepsilon\,,
\end{equation}
then, for all $m<m'$,
\begin{equation}
\label{ChiDifferenceBound}
\int (\overline{\chi}_m(x) - \overline{\chi}_{m'}(x))^2 \,dP \leq \frac{2}{m}\bigg(1-\frac{m}{m'}\bigg) + 4\varepsilon\bigg(1 - \frac{m}{m'}\bigg)^{\!\!2}\,.
\end{equation}
\null\hfill $\square$
\end{lem}

Using this lemma, we can prove also the following one, which is a generalization for $K$ random variables of an additional result by Koml\'{o}s \cite[Lemma 3]{komlos1}.

\begin{lem}
\label{KomlosLemma2}
Consider any fixed sequence of $K$ symbols $\mathbf{x}=(x_1,\dots, x_{k}, x_{k+1},\dots,x_K) \in \mathcal{X}^K$, and consider a sequence $\mathbf{x}' = (x_1, \dots, x_{k+1}, x_{k}, \dots, x_K)$ obtained from $\mathbf{x}$ by swapping any two adjacent symbols. Suppose that for all $x \in \mathcal{X}$, the functions $\chi_m(x)$ satisfy the condition in \eqref{KomlosLemma1Cond} with the same $\varepsilon$. If for all ordered subsets of $K$ random variables $\{X_{m_1},X_{m_2},\dots,X_{m_K}\} \subset \mathcal{M}$, $m_i < m_j\,$ for $i<j$,
\begin{equation}
\label{KomlosLemma2H1}
\bigg\lvert\int \chi_{m_1}(x_1)\cdots\chi_{m_k}(x_k)\,\chi_{m_{k+1}}(x_{k+1})\cdots\chi_{m_K}(x_K)dP - r_{\mathbf{x}}\,\bigg\rvert \leq \delta
\end{equation}
and
\begin{equation}
\label{KomlosLemma2H2}
\bigg\lvert\int \chi_{m_1}(x_1)\cdots\chi_{m_k}(x_{k+1})\,\chi_{m_{k+1}}(x_k)\cdots\chi_{m_K}(x_K)\,dP - r_{\mathbf{x}'}\,\bigg\rvert \leq \delta\,,
\end{equation}
then
\begin{equation}
\label{KomlosLemma2Result}
\lvert\, r_{\mathbf{x}} - r_{\mathbf{x}'}\rvert \leq 4K\sqrt{\frac{2K}{M}} + 8K\sqrt{\varepsilon} + 2\delta.
\end{equation}
\null\hfill $\square$
\end{lem}

\begin{IEEEproof}
The proof is similar to the original by Koml\'{o}s, with minor adjustments. First of all, for each $x \in \mathcal{X}$, we split the sequence $\chi_1(x),\dots,\chi_{M}(x)$ into $K$ consecutive blocks and we define the averages
\begin{equation}
\label{AtDef}
A_l(x) \triangleq \frac{K}{M} \big(\chi_{(l-1)M/K + 1}(x)+\chi_{(l-1)M/K + 2}(x)\dots +\chi_{lM/K}(x)\big)
\end{equation}
for every $1 \leq l \leq K$.
Then, by \eqref{KomlosLemma2H1} and the triangle inequality we have that for all $M/K < m_2 < \dots < m_K$,
\begin{equation*}
\bigg\lvert\int A_1(x_1)\,\chi_{m_2}(x_2)\cdots\chi_{m_K}(x_K)\,dP - r_{\mathbf{x}}\,\bigg\rvert \leq \delta\,.
\end{equation*}
Furthermore, for all $2M/K < m_3 < \dots < m_K$,
\begin{equation*}
\bigg\lvert\int A_1(x_1)\,A_2(x_2)\,\chi_{m_3}(x_3)\cdots\chi_{m_K}(x_K)\,dP - r_{\mathbf{x}}\,\bigg\rvert \leq \delta\,.
\end{equation*}
Proceeding in the same way we obtain
\begin{equation}
\label{Bound1}
\bigg\lvert\int A_1(x_1)\cdots A_k(x_k)\,A_{k+1}(x_{k+1})\cdots A_K(x_K)\,dP - r_{\mathbf{x}}\,\bigg\rvert \leq \delta\,.
\end{equation}
In the same way, using \eqref{KomlosLemma2H2} we get
\begin{equation}
\label{Bound2}
\bigg\lvert\int\!\! A_1(x_1)\cdots A_k(x_{k+1})A_{k+1}(x_k)\cdots A_K(x_K)\,dP - r_{\mathbf{x}'}\,\bigg\rvert \leq \delta\,.
\end{equation}
Next, using the Cauchy-Schwarz inequality we have that
\begin{align}
\label{IntBound1}
\bigg\lvert\int A_1(x_1)\cdots \big[A_k(x_k)\,&A_{k+1}(x_{k+1}) -A_k(x_{k+1})\,A_{k+1}(x_k)\big]\cdots A_K(x_K)\,dP\,\bigg\rvert \\
&\leq \big\lVert\, A_k(x_k)\,A_{k+1}(x_{k+1})-A_k(x_{k+1})\,A_{k+1}(x_k)\,\big\rVert \notag\\
&= \big\lVert\, A_k(x_k)\,A_{k+1}(x_{k+1})-A_{k+1}(x_k)\,A_{k+1}(x_{k+1}) \\
&\hspace{8em}+A_{k+1}(x_k)\,A_{k+1}(x_{k+1})-A_k(x_{k+1})\,A_{k+1}(x_k)\,\big\rVert \notag\\
&\leq \big\lVert A_k(x_k)-A_{k+1}(x_k)\big\rVert + \big\lVert A_{k+1}(x_{k+1})-A_k(x_{k+1})\big\rVert\,,
\end{align}
where we used the fact that $0 \leq A_l(x) \leq 1$ over the whole sample space $\Omega$, for every $1\leq l \leq K$ and $x \in \mathcal{X}$.
Furthermore, using \eqref{AverageDef}, one can verify that, for any $1\leq k \leq K-1$ and $x\in\mathcal{X}$,
\begin{equation}
(k+1)\left(\overline{\chi}_{kM/K}(x) - \overline{\chi}_{(k+1)M/K}(x)\right) = \frac{K}{M}\left(\frac{1}{k}\sum_{i=1}^{kM/K} \chi_{i}(x) - \sum_{i=kM/K + 1}^{(k+1)M/K} \chi_{i}(x)\right)
\end{equation}
and
\begin{equation}
(k-1)\left(\overline{\chi}_{kM/K}(x) - \overline{\chi}_{(k-1)M/K}(x)\right) = \frac{K}{M}\left(-\frac{1}{k}\sum_{i=1}^{kM/K} \chi_{i}(x) + \sum_{i=(k-1)M/K + 1}^{kM/K} \chi_{i}(x)\right)
\end{equation}
and therefore, using definition \eqref{AtDef}, one can write
\begin{equation}
\label{AkDifference}
A_k(x) - A_{k+1}(x) = (k+1)\left(\overline{\chi}_{kM/K}(x)- \overline{\chi}_{(k+1)M/K}(x)\right) + (k-1)\left(\overline{\chi}_{kM/K}(x)- \overline{\chi}_{(k-1)M/K}(x)\right).
\end{equation}
Then, one can use equation \eqref{ChiDifferenceBound} and the fact that $k\leq K-1$ to bound the norm of the terms in \eqref{AkDifference}, obtaining
\begin{equation}
\big\lVert\,A_k(x) - A_{k+1}(x)\,\big\rVert \leq 2K\sqrt{\frac{2K}{M}} + 4K\sqrt{\varepsilon}, \qquad 1\leq k\leq K-1.
\end{equation}
Therefore, the last line of equation \eqref{IntBound1} can be upper bounded by $4K\sqrt{\frac{2K}{M}} + 8K\sqrt{\varepsilon}$.
Finally, this and equations \eqref{Bound1} and \eqref{Bound2} lead to \eqref{KomlosLemma2Result}.
\end{IEEEproof}

Notice that if we have a set of random variables $\mathcal{M}$ for which the hypotheses of Lemma \ref{KomlosLemma2} hold for \emph{any} sequence of $K$ symbols $\mathbf{x}$, then we can bound $\lvert r_{\mathbf{x}} - r_{\mathbf{x}'}\rvert$ for \emph{any permutation} $\mathbf{x}'$ of $\mathbf{x}$, since any permutation of $\mathbf{x}$ can be obtained as a succession of adjacent elements swaps. Since the number of swaps is lower than $K^2/2$, we obtain the bound
\[
\lvert\, r_{\mathbf{x}} - r_{\mathbf{x}'}\rvert \leq 2K^3\sqrt{\frac{2K}{M}} + 4K^3\sqrt{\varepsilon} + K^2\delta\,.
\]

We can now link this result on random variables to codes through the natural association between codewords and random variables. Consider the probability space made of the sample space $\Omega = \{1,2,\dots,n\}$, the $\sigma$-algebra $\mathsf{P}(\Omega)$ (the power set of $\Omega$), and the probability measure $P$ such that $P(i)=1/n$ for every $i\in\Omega$. Then we can associate to each codeword $\mathbf{x}_m$ a random variable $X_m$ that takes values $X_m(i) = x_{m,i}$ for every $i\in\Omega$, where $x_{m,i}$ is the symbol in the $i$-th coordinate of $\mathbf{x}_m$. Hence, if, for a generic $K$, $q_{\mathbf{m}}(\mathbf{x})$ is the fraction of times the sequence of symbols $\mathbf{x}=(x_1,x_2,\dots,x_K)$ appears at the same coordinate in the group of codewords $\mathbf{m}=(\mathbf{x}_{m_1},\mathbf{x}_{m_2},\dots,\mathbf{x}_{m_K})$ -- i.e., the joint type of the code $\mathbf{m}$, -- then
\[
q_{\mathbf{m}}(\mathbf{x})=P(X_{m_1}=x_1,X_{m_2}=x_2,\dots,X_{m_K}=x_K).
\]
In such a way, we can combine Lemma \ref{KomlosLemma1} and Lemma \ref{KomlosLemma2} and the remark immediately afterwards to obtain the following result on codes.
\begin{lem}
\label{LemmaCodes}
Consider a code $\mathcal{C}$ with $M$ codewords $\{\mathbf{x}_1,\mathbf{x}_2,\dots,\mathbf{x}_M\}$ of length $n$. If for each $x \in \mathcal{X}$ there exists a number $r_x$ such that for all $m<m'$,
\begin{equation}
\label{LemmaCodesCondition1}
\big\lvert\, q_{m,m'}(x,x)-r_x\,\big\rvert \leq \varepsilon\,,
\end{equation}
and if for each sequence of $K$ symbols $\mathbf{x}=(x_1,x_2,\dots,x_K) \in \mathcal{X}^K$ there exists a number $r_{\mathbf{x}}$ such that for all ordered subsets of $K$ codewords $\mathbf{m}=\{\mathbf{x}_{m_1},\mathbf{x}_{m_2},\dots,\mathbf{x}_{m_K}\}$, $m_i < m_j\,$ for $i<j$, 
\begin{equation}
\label{LemmaCodesCondition2}
\big\lvert\, q_{\mathbf{m}}(\mathbf{x})-r_{\mathbf{x}}\,\big\rvert \leq \delta\,,
\end{equation}
then for any permutation $\mathbf{x}'$ of $\mathbf{x}$,
\begin{equation}
\big\lvert\, q_{\mathbf{m}}(\mathbf{x}) - q_{\mathbf{m}}(\mathbf{x}')\,\big\rvert \leq 2K^3\sqrt{\frac{2K}{M}} + 4K^3\sqrt{\varepsilon} + (K^2+2)\delta\,.
\end{equation}
\null\hfill $\square$
\end{lem}

We now show, using Ramsey's theorem for hypergraphs, that from a code large enough we can always extract a subcode that satisfies the conditions of Lemma \ref{LemmaCodes}, whose size grows unbounded as the size of the original code tends to infinity.

\begin{thm}[Ramsey's theorem for hypergraphs \cite{diestel1}]
For any positive integers $K$, $C$ and $M$, there exists a positive integer $M_0 \geq M$ such that any complete $K$-hypergraph with at least $M_0$ vertices, edge-colored with $C$ colors in any way, contains a complete monochromatic subgraph with at least $M$ vertices.
\end{thm}

\begin{thm}
\label{KomlosTheorem}
For any integer $t>0$ there exists a positive integer $M_0$ such that from any code $\mathcal{C}$ with $M\geq M_0$ codewords a subcode $\mathcal{C'} \subset \mathcal{C}$ can be extracted with $M'$ codewords $\{\mathbf{x}_1,\mathbf{x}_2,\dots,\mathbf{x}_{M'}\}$, with $M' \to \infty$ as $M \to \infty$, such that for any subset of $K$ codewords $\mathbf{m} \subset \mathcal{C'}$, for any sequence of input symbols $\mathbf{x}$ and any of its permutations $\mathbf{x}'$, 
\begin{equation}
\label{KomlosTheoremResult}
\big\lvert\, q_{\mathbf{m}}(\mathbf{x}) - q_{\mathbf{m}}(\mathbf{x}')\,\big\rvert \leq \Delta(M',t)\,,
\end{equation}
where $\Delta(M',t) \triangleq 2K^3\sqrt{\frac{2K}{M'}} + 4K^3\sqrt{\frac{\lvert\mathcal{X}\rvert^{K-2}}{2t}} + \frac{K^2+2}{2t}$.\hfill $\square$
\end{thm}
\begin{IEEEproof}
Consider a complete hypergraph with $M$ vertices, where each vertex is associated with a different codeword of $\mathcal{C}$, and each edge is an ordered subset of $K$ vertices --- i.e., of $K$ ordered codewords. We color each edge of the graph with a \emph{vector-color} with $\lvert\mathcal{X}\rvert^K$ components, each corresponding to one of the sequences of $K$ input symbols $\mathbf{x}$. For each component we define $t$ possible colors, corresponding to the $t$ equal-length subintervals of the interval $[0,1]$. To each edge $\mathbf{m} \in \mathcal{C}$ (an ordered subset of $K$ codewords) we assign as a color to each component $\mathbf{x}$ the subinterval of $[0,1]$ that contains the value of $q_{\mathbf{m}}(\mathbf{x})$. 

By Ramsey's theorem for hypergraphs (see for example \cite{diestel1}), if $M \geq K$, we can always extract a complete \emph{monochromatic} subgraph whose size grows unbounded as $M$ goes to infinity.

If we choose the vertices of this subgraph as our subcode $\mathcal{C}'$, the graph being monochromatic means that for any $\mathbf{x} \in \mathcal{X}^K$, $q_{\mathbf{m}}(\mathbf{x})$ is in the same subinterval of $[0,1]$ for all ordered $\mathbf{m} \in \mathcal{C}'$; this means that $\mathcal{C}'$ meets condition \eqref{LemmaCodesCondition2} with the midpoint of the subinterval as $r_{\mathbf{x}}$, and $\delta$ equal to half the length of the subinterval, i.e., $\delta = 1/(2t)$. 

We now show that if the size of $\mathcal{C}'$ is greater than $2K-2$, then the subcode of $\mathcal{C'}$ obtained removing the last $K-2$ codewords also meets condition \eqref{LemmaCodesCondition1}. In fact, consider all the ordered edges of $\mathcal{C'}$ such that the last $K-2$ codewords are fixed as the last $K-2$ codewords of $\mathcal{C}'$, and the first two are taken from all the ordered pairs $(m,m')$ of the other codewords. Since all these edges have the same color, it follows that for each pair $(m,m')$, $q_{m,m'}(x,x) = \sum_{\mathbf{x}} q_{\widetilde{\mathbf{m}}}(\mathbf{x})$, where $\widetilde{\mathbf{m}}$ is the concatenation of $(m,m')$ and the last $K-2$ codewords of $\mathcal{C'}$, and the sum is over all $\mathbf{x} \in \mathcal{X}^K$ such that $x_1=x_2=x$. Since the $\widetilde{\mathbf{m}}$'s have the same vector-color for all $(m,m')$, we have that $\big\lvert\,q_{\widetilde{\mathbf{m}}}(\mathbf{x}) - r_{\mathbf{x}}\,\big\rvert \leq\delta$ for all $\widetilde{\mathbf{m}}$.
If we define $r \triangleq \sum_{\mathbf{x}} r_{\mathbf{x}}$, where the sum is again over all $\mathbf{x}$ with $x_1=x_2=x$, then it follows that for any $x \in \mathcal{X}$, for all $(m,m')$,
\begin{equation}
\big\lvert\,q_{m,m'}(x,x)-r\,\big\rvert \leq \sum_{\mathbf{x}}\big\lvert\,q_{\widetilde{\mathbf{m}}}(\mathbf{x})-r_{\mathbf{x}}\,\big\rvert \leq \lvert\mathcal{X}\rvert^{K-2}\,\delta\,,
\end{equation}
so that $\mathcal{C}'$ without the last $K-2$ codewords meets condition \eqref{LemmaCodesCondition1} with $\varepsilon = \lvert\mathcal{X}\rvert^{K-2}\,\delta =  \lvert\mathcal{X}\rvert^{K-2}/(2t)$, in addition to condition \eqref{LemmaCodesCondition2}, which is inherited from $\mathcal{C}'$ provided that the latter has at least $K$ codewords other than the last $K-2$. Ramsey's theorem satisfies this last condition provided that the starting code $\mathcal{C}$ is greater than a certain finite number $M_0$ that depends on $t$. Finally, equation \eqref{KomlosTheoremResult} follows from Lemma \ref{LemmaCodes} with $\delta = 1/(2t)$ and $\varepsilon = \lvert\mathcal{X}\rvert^{K-2}/(2t)$.
\end{IEEEproof}

For the case $L=1$, Theorem 3.4 can be restated equivalently in terms of random variables following the original formulation by Koml\'{o}s, as follows.
\begin{thm}[Koml\'{o}s \cite{komlos1}]
\label{KomlosTheoremRV}
For any integer $t>0$ we can extract from any set of random variables $\mathcal{M}=\big\{X_1,X_2,\dots,X_M\big\}$ taking values in a finite alphabet $\mathcal{X}$, a subset $\mathcal{M'} \subset \mathcal{M}$ of $M'$ random variables, with $M' \to \infty$ as $M \to \infty$, such that for any pair of random variables $X_m, X_{m'} \subset \mathcal{M'}$, $m<m'$, for any pair of values $x,x' \in \mathcal{X}$,
\begin{equation}
\label{KomlosTheoremResult2}
\big\lvert\, P(X_m=x, X_{m'}=x') - P(X_m=x', X_{m'}=x)\,\big\rvert \leq \Delta(M',t)\,,
\end{equation}
where $\Delta(M',t) \to 0$ as $M', t \to \infty$.\hfill $\square$
\end{thm}

It is worth noting that in \cite{komlos1}, Koml\'{o}s presents as its main result the following weaker theorem on the symmetry of a pair of random variables.
\begin{thm}[Koml\'{o}s \cite{komlos1}]
\label{KomlosTheoremRV2}
From any set of random variables $\mathcal{M}=\big\{X_1,X_2,\dots,X_M\big\}$ taking values in a finite alphabet $\mathcal{X}$, there exists a pair of random variables $X_m$ and $X_{m'}$ such that
\begin{equation}
\label{KomlosTheoremResult3}
\big\lvert\, P(X_m > X_{m'}) - P(X_m < X_{m'})\,\big\rvert \to 0 \quad \text{as } M\to\infty.
\end{equation}
\null\hfill $\square$
\end{thm}

It can be shown that this same result (published by Koml\'{o}s in 1990) can also be obtained following step-by-step Berlekamp's proof of the bound on the zero-rate reliability function \cite{sgb2} (originally published in his PhD dissertation in 1964), the only change being the substitution of $\mu(s)$ with a different function. This fact shows that the proof presented here and Berlekamp's are much more deeply connected than one would think.

\section{Bound on $E_L(0)$ for $L$-list decoding}
\label{sec:boundE_L(0)}

Gallager \cite{gallager2} derived a well-known lower bound for the reliability function, the \emph{expurgated bound}, which can be easily generalized to list decoding using the same reasoning. This bound at rate $R=0$ assumes the form
\begin{equation}
\label{ExBound}
E_L(0^+) \geq \max_{Q \in \mathcal{P}(\mathcal{X})}\left[-\sum_{\mathbf{x} \in \mathcal{X}^{L+1}} Q(x_1)\cdots Q(x_{L+1})\log \sum_{y \in \mathcal{Y}} \sqrt[L+1]{P(y|x_1)\cdots P(y|x_{L+1})}\right].
\end{equation}
We can use Theorem \ref{KomlosTheorem} with $K=L+1$ to obtain an upper bound on $E_L(0)$. In fact, starting from any code $\mathcal{C}$ with $M$ codewords of length $n$ we can extract the subcode $\mathcal{C}'$ with $M'$ codewords of lengh $n$ indicated by Theorem \ref{KomlosTheorem}. Since $\mathcal{C}' \subset \mathcal{C}$, we have that, according to equation \eqref{Dmin}, $D_{\min}(\mathcal{C}) \leq D_{\min}(\mathcal{C}')$.
Moreover, for any subset of $L+1$ codewords $\mathbf{m} \in \mathcal{C}'$ we have
\begin{align}
D_{\mathbf{m}} &= \max_{\bm{\alpha}} \sum_{\mathbf{x} \in \mathcal{X}^{L+1}} q_{\mathbf{m}}(\mathbf{x}) \,\mu_{\mathbf{x}}(\bm{\alpha}) \notag \\
	&=\max_{\bm{\alpha}} \sum_{\mathbf{x} \in \mathcal{X}_*^{L+1}} \sum_{\mathbf{x}' \in S(\mathbf{x})} q_{\mathbf{m}}(\mathbf{x}') \,\mu_{\mathbf{x'}}(\bm{\alpha})\,,
\end{align}
where $\mu_{\mathbf{x}}(\bm{\alpha})$ is defined in \eqref{MuX}, $\mathcal{X}_*^{L+1}$ is the set of of all sequences of $L+1$ input symbols $(x_1,x_2,\dots,x_{L+1})$ such that $x_1\leq x_2 \leq \cdots\leq x_{L+1}$, and $S(\mathbf{x})$ is the set of all permutations of $\mathbf{x}$. Then, by equation \eqref{KomlosTheoremResult},
\begin{align}
D_{\mathbf{m}} &\leq \max_{\bm{\alpha}} \sum_{\mathbf{x} \in \mathcal{X}_*^{L+1}} \sum_{\mathbf{x}' \in S(\mathbf{x})} \big(q_{\mathbf{m}}(\mathbf{x})+\Delta(M',t)\big) \mu_{\mathbf{x'}}(\bm{\alpha}) \notag\\
	&\leq \bigg(\max_{\bm{\alpha}} \sum_{\mathbf{x} \in \mathcal{X}_*^{L+1}}\!\! q_{\mathbf{m}}(\mathbf{x}) \!\sum_{\mathbf{x}' \in S(\mathbf{x})}\!\!\mu_{\mathbf{x'}}(\bm{\alpha})\bigg) +\Delta(M',t)\, \max_{\bm{\alpha}}\sum_{\mathbf{x} \in \mathcal{X}^{L+1}}\mu_{\mathbf{x}}(\bm{\alpha})\,. \label{DmBound1}
\end{align}
Since for any $\mathbf{x} \in \mathcal{X}_*^{L+1}$, the function $\sum_{\mathbf{x}' \in S(\mathbf{x})}\mu_{\mathbf{x'}}(\bm{\alpha})$ is symmetric (that is, invariant to permutations of its arguments) and concave, it follows that the term in parentheses is maximized for $\widetilde{\bm{\alpha}} \triangleq \big(\frac{1}{L+1},\dots,\frac{1}{L+1}\big)$, and therefore we can write
\begin{align}
\max_{\bm{\alpha}} \sum_{\mathbf{x} \in \mathcal{X}_*^{L+1}}\!\! q_{\mathbf{m}}(\mathbf{x}) \!\sum_{\mathbf{x}' \in S(\mathbf{x})}\!\!\mu_{\mathbf{x'}}(\bm{\alpha}) &= \sum_{\mathbf{x} \in \mathcal{X}_*^{L+1}} \sum_{\mathbf{x}' \in S(\mathbf{x})} q_{\mathbf{m}}(\mathbf{x}) \,\mu_{\mathbf{x'}}(\widetilde{\bm{\alpha}})\\
&\leq\sum_{\mathbf{x} \in \mathcal{X}_*^{L+1}} \sum_{\mathbf{x}' \in S(\mathbf{x})}\big(q_{\mathbf{m}}(\mathbf{x}')+\Delta(M',t)\big)\mu_{\mathbf{x'}}(\widetilde{\bm{\alpha}})\,,
\end{align}
where the inequality in the second line is again due to \eqref{KomlosTheoremResult}. Hence, this and equation \eqref{DmBound1} lead to
\begin{equation}
D_{\mathbf{m}} \leq \sum_{\mathbf{x} \in \mathcal{X}^{L+1}} q_{\mathbf{m}}(\mathbf{x}) \,\mu_{\mathbf{x}}(\widetilde{\bm{\alpha}}) +C\,\Delta(M',t)\,,
\end{equation}
where we defined the finite positive quantity $C \triangleq \max_{\bm{\alpha}}\sum_{\mathbf{x} \in \mathcal{X}^{L+1}}\big(\mu_{\mathbf{x}}(\bm{\alpha})+\mu_{\mathbf{x}}(\widetilde{\bm{\alpha}})\big)$.
Next, since $D_{\min}(\mathcal{C'})$ is lower than or equal to the average of $D_{\mathbf{m}}$ over all subsets of $L+1$ codewords $\mathbf{m} \subset \mathcal{C'}$, it follows that
\begin{align}
D_{\min}(\mathcal{C}') &\leq \frac{1}{M'(M'-1)\cdots (M'-L)} \sum_{\mathbf{m}} D_{\mathbf{m}} \notag\\
	&\leq C\,\Delta(M',t) + \frac{1}{(M'-L)^{L+1}} \sum_{\mathbf{m}}\sum_{\mathbf{x} \in \mathcal{X}^{L+1}} q_{\mathbf{m}}(\mathbf{x}) \,\mu_{\mathbf{x}}(\widetilde{\bm{\alpha}})\,. \label{DmBound2}
\end{align}
The double sum can be computed on a column-by-column basis, as in the derivation of the Plotkin bound. Letting $M_c(x)$ be the number of times the input symbol $x$ appears in the column $c$ over all the codewords of $\mathcal{C}'$, we can write
\begin{align}
\sum_{\mathbf{m}} \sum_{\mathbf{x} \in \mathcal{X}^{L+1}} q_{\mathbf{m}}(\mathbf{x}) \,\mu_{\mathbf{x}}(\widetilde{\bm{\alpha}}) &= \frac{1}{n}\sum_{c=1}^n \sum_{\mathbf{x} \in \mathcal{X}^{L+1}} \!\!M_c(x_1)\,\cdots M_c(x_{L+1}) \,\mu_{\mathbf{x}}(\widetilde{\bm{\alpha}})\\
	&= \frac{M'^{L+1}}{n}\sum_{c=1}^n \sum_{\mathbf{x} \in \mathcal{X}^{L+1}} \frac{M_c(x_1)}{M'}\cdots \frac{M_c(x_{L+1})}{M'} \mu_{\mathbf{x}}(\widetilde{\bm{\alpha}}) \label{DmBoundMc}\\
	&\leq M'^{L+1} \max_{Q \in \mathcal{P}(\mathcal{X})}\sum_{\mathbf{x} \in \mathcal{X}^{L+1}} Q(x_1)\cdots Q(x_{L+1})\,\mu_{\mathbf{x}}(\widetilde{\bm{\alpha}})\,. \label{DmBound3}
\end{align}
Then, if we put equation \eqref{DmBound3} into \eqref{DmBound2} we get
\begin{equation}
D_{\min}(\mathcal{C}') \leq C\,\Delta(M',t) + \bigg(\frac{M'}{M'-L}\bigg)^{L+1} \max_{Q \in \mathcal{P}(\mathcal{X})}\sum_{\mathbf{x} \in \mathcal{X}^{L+1}} Q(x_1)\cdots Q(x_{L+1})\,\mu_{\mathbf{x}}(\widetilde{\bm{\alpha}})\,.
\end{equation}
Notice that the obtained bound is independent of the actual code $\mathcal{C}$. We can now take the limits $n\to\infty$ and $t\to \infty$ for any code $\mathcal{C}$ of any rate $R>0$ to get
\begin{equation}
D_{\min}(\mathcal{C}) \leq \max_{Q \in \mathcal{P}(\mathcal{X})}\sum_{\mathbf{x} \in \mathcal{X}^{L+1}} Q(x_1)\cdots Q(x_{L+1})\,\mu_{\mathbf{x}}(\widetilde{\bm{\alpha}})\,,
\end{equation}
since for any $R>0$, taking $n\to\infty$ implies that $M\to\infty$, and therefore $M'\to\infty$.
Finally, by equation \eqref{PeMLower}, since the bound holds for any code $\mathcal{C}$ and any rate $R>0$, we get the following upper bound on the reliability function of the channel under consideration at rate $R \to 0$:
\begin{equation}
\label{FinalBound}
E_L(0^+) \leq \max_{Q \in \mathcal{P}(\mathcal{X})}\sum_{\mathbf{x} \in \mathcal{X}^{L+1}} Q(x_1)\cdots Q(x_{L+1})\,\mu_{\mathbf{x}}(\widetilde{\bm{\alpha}})\,.
\end{equation}
Using the definition of $\mu_{\mathbf{x}}(\tilde{\bm{\alpha}})$ we can see that the upper bound is exactly equal to the expurgated lower bound \eqref{ExBound}, proving that this value is precisely $E_L(0^+)$.

\section{Blahut's low rate bound for constant composition codes}
\label{sec:Blahut}
We analyze in this Section an upper bound on the reliability function of discrete memoryless channels at low rates proposed in \cite{blahut}, for constant composition codes, that is, codes for which every codeword has the same composition or type. The bound is stated there for $L=1$  and for the specific class of \emph{nonnegative-definite channels} originally studied by Jelinek \cite{jelinek1}. However, as already discussed in \cite[Sec. VI.A]{dalai-2015}, there is a major gap in the proof since it is erroneously based on the assumption that the Bhattacharyya distance between two codewords with identical composition can be used as an upper bound on the error exponent in a binary hypothesis test between those two codewords. 

In this Section we show that the proof can be fixed in a rather simple way using the approach discussed in the previous sections, and can even be extended to $L>1$ and to any discrete memoryless channel. In fact, the restriction to a specific class of channels made in \cite{blahut} is motivated only by the requirement that a certain quadratic form be concave (see \cite{dalai-2015} for details). By replacing this quadratic form with its upper concave envelope, one can prove the bound for any channel, and this bound turns out to be tight at $R=0$ for constant composition codes, for which the true value of the reliability for $L=1$ already takes a different form with respect to \eqref{ExBound} (see \cite[Prob. 10.22]{csiszar1}).


We define here quantities for constant composition codes analogous to the ones considered in the previous part of the paper. Let $P_e(L,R,n,Q)$ be the smallest probability of error for $L$-list decoding over all codes with rate at least $R$, block length $n$ and codewords with composition $Q$. Then let 
\begin{equation}
\label{ELDefQ}
E_L(R,Q) \triangleq \limsup_{n \to \infty} -\frac{\log P_e(L,R,n,Q_n)}{n}\,.
\end{equation}
where $Q$ is a general distribution over $\mathcal{X}$ and the $\limsup$ is over all sequences of codes with constant compositions $Q_n$ such that $\lim_{n\to\infty}Q_n=Q$. 

The idea of the bound is essentially the same used for the Elias bound on the minimum distance of binary codes (see \cite{dalai-2015} for a detailed discussion). Consider any code $\mathcal{C}$ of rate $R>0$, blocklength $n$ and constant composition $Q$. Instead of extracting directly a symmetric subcode using the results of Section III, we first extract another subcode $\mathcal{C}'$ for which all the codewords have a fixed conditional type with respect to a given auxiliary sequence $\mathbf{a}$. We choose this auxiliary sequence by means of the following lemma.


\begin{lem}[See \cite{dalai-2015}]
\label{le:subcode}
Let $\mathcal{C}$ be any constant composition code with codewords of composition $Q$. Let $\mathcal{A}$ be any auxiliary set, $F$ be a type for sequences in $\mathcal{A}^n$ and $V$ a conditional type for sequences in $\mathcal{X}^n$ given sequences in $\mathcal{A}^n$ of type $F$, such that the resulting type for sequences in $\mathcal{X}^n$ is $Q$, that is, such that
\begin{equation}
\sum_{a\in\mathcal{A}} F(a)V(x|a) = Q(x)
\end{equation}
for every $x\in\mathcal{X}$. We will denote this condition by $FV=Q$. Then, there is a sequence $\mathbf{a}\in\mathcal{A}^n$ of type $F$ and a subcode $\mathcal{C}'$  with $\lvert\mathcal{C}'\rvert \geq \lvert\mathcal{C}\rvert\cdot e^{-n(I(F,V)+o(1))}$, such that all its codewords have conditional type $V$ given $\mathbf{a}$, where

\begin{equation}
I(F,V) \triangleq \sum_{a\in\mathcal{A}} \sum_{x\in\mathcal{X}} F(a)V(x|a)\log \frac{V(x|a)}{\sum_{a' \in\mathcal{A}} F(a')V(x|a')}\,.
\end{equation}
\null\hfill $\square$
\end{lem}


Consider now any $F$ and $V$ such that $FV=Q$ and $I(F,V) < R$. Then, by Lemma \ref{le:subcode}, we can extract a subcode $\mathcal{C}'$ from $\mathcal{C}$ with $\lvert\mathcal{C}'\rvert \geq e^{n(R-I(F,V)+o(1))}$ codewords. Notice that $\lvert\mathcal{C}'\rvert \to \infty$ as $n\to\infty$. Next, for $n$ large enough, we can use Theorem \ref{KomlosTheoremRV} to extract a symmetric subcode $\mathcal{C}''$ from $\mathcal{C}'$ such that, again, $\lvert\mathcal{C}''\rvert \to \infty$ as $n\to\infty$.

Let now $M'' = \lvert\mathcal{C}''\rvert$. Following the same steps as in Section IV up until equation \eqref{DmBoundMc}, we get

\begin{equation}
D_{\min}(\mathcal{C}) \leq C\,\Delta(M'',t) + \bigg(\frac{M''}{M''-L}\bigg)^{L+1}\frac{1}{n}\sum_{c=1}^n \sum_{\mathbf{x} \in \mathcal{X}^{L+1}} \frac{M''_c(x_1)}{M''}\cdots \frac{M''_c(x_{L+1})}{M''} \mu_{\mathbf{x}}(\widetilde{\bm{\alpha}})\,,
\end{equation}
where $M''_c(x)$ is the number of times the input symbol $x$ appears in coordinate $c$ over all the codewords of $\mathcal{C}''$. Define for notational convenience the probability distributions $\lambda_c$ as
$$
\lambda_c(x)=\frac{M_c''(x)}{M''}\,.
$$
Since $M''\to\infty$ as $n\to\infty$, we can write
\begin{equation}
\label{DminLambda}
D_{\min}(\mathcal{C}) \leq \frac{1}{n}\sum_{c=1}^n \sum_{\mathbf{x} \in \mathcal{X}^{L+1}} \lambda_c(x_1)\cdots \lambda_c(x_{L+1}) \mu_{\mathbf{x}}(\widetilde{\bm{\alpha}}) + o(1).
\end{equation}
The quantity $\mu(\widetilde{\bm{\alpha}})$ is the equivalent of the Bhattacharyya distance for list $L$. The error in Blahut's proof was that he used this distance directly for $\mathcal{C}'$, but since $\mathcal{C}'$ is not a symmetric subcode (only a constant composition one), his bound on the minimum distance does not translate into a valid bound on the probability of error. This is instead true in our case since, just like in the zero-rate case, the symmetry of $\mathcal{C}''$ allows us to substitute $\max_{\bm{\alpha}}\mu(\bm{\alpha})$ with $\mu(\widetilde{\bm{\alpha}})$ with asymptotically negligible error.

Next, notice that the probability distributions $\lambda_c$ satisfy the conditions
\begin{equation}
\label{LambdaConds}
\frac{1}{nF(a)}\sum_{c=1}^n\mathbbm{1}_{\{a_c=a\}}\lambda_c(x)=V(x|a)\,,
\end{equation}
for every $x\in\mathcal{X}$ and every $a\in\mathcal{A}$ such that $F(a) > 0$, where $\mathbbm{1}_{\{a_c=a\}}$ equals $1$ if $\mathbf{a}$ has symbol $a$ in the $c$-th coordinate, and $0$ otherwise. Furthermore, one can rewrite \eqref{DminLambda} as
\begin{equation}
\label{DminLambda2}
D_{\min}(\mathcal{C}) \leq \frac{1}{n} \sum_{a} \sum_{c=1}^n\mathbbm{1}_{\{a_c=a\}}\sum_{\mathbf{x} \in \mathcal{X}^{L+1}} \lambda_c(x_1)\cdots \lambda_c(x_{L+1}) \mu_{\mathbf{x}}(\widetilde{\bm{\alpha}}) + o(1)\,,
\end{equation}
where the first sum is over all $a\in\mathcal{A}$ such that $F(a)>0$. Let now
\begin{equation}
\mathfrak{C}\left[\sum_{\mathbf{x} \in \mathcal{X}^{L+1}}\lambda_c(x_1)\cdots \lambda_c(x_{L+1}) \mu_{\mathbf{x}}(\widetilde{\bm{\alpha}})\right]
\end{equation}
be the upper concave envelope of $\sum_{\mathbf{x} \in \mathcal{X}^{L+1}}\lambda_c(x_1)\cdots \lambda_c(x_{L+1}) \mu_{\mathbf{x}}(\widetilde{\bm{\alpha}})$ as a function of $\lambda_c$. We can further upper bound \eqref{DminLambda2} by
\begin{align}
D_{\min}(\mathcal{C}) &\leq \frac{1}{n} \sum_{a} \sum_{c=1}^n\mathbbm{1}_{\{a_c=a\}}\mathfrak{C}\left[\sum_{\mathbf{x} \in \mathcal{X}^{L+1}} \lambda_c(x_1)\cdots \lambda_c(x_{L+1}) \mu_{\mathbf{x}}(\widetilde{\bm{\alpha}})\right] + o(1) \\
	&= \sum_{a} F(a)\sum_{c=1}^n\frac{\mathbbm{1}_{\{a_c=a\}}}{nF(a)}\mathfrak{C}\left[\sum_{\mathbf{x} \in \mathcal{X}^{L+1}} \lambda_c(x_1)\cdots \lambda_c(x_{L+1}) \mu_{\mathbf{x}}(\widetilde{\bm{\alpha}})\right] + o(1)\,.
\end{align}
Notice that $\sum_c \frac{\mathbbm{1}_{\{a_c=a\}}}{nF(a)} = 1$. Hence, we can use Jensen's inequality to get
\begin{equation}
D_{\min}(\mathcal{C}) \leq \sum_{a} F(a)\,\mathfrak{C}\left[\sum_{\mathbf{x} \in \mathcal{X}^{L+1}} V(x_1|a)\cdots V(x_{L+1}|a) \mu_{\mathbf{x}}(\widetilde{\bm{\alpha}})\right] + o(1)\,,
\end{equation}
where we also used \eqref{LambdaConds}.
Since this bound holds for any code of rate $R$ and $n$ large enough, equations \eqref{PeMLower} and \eqref{ELDefQ} lead to the following bound on the $L$-list reliability function for codes of rate $R$ and constant composition $Q$:
\begin{equation}
\label{ELBoundQ}
E_L(R,Q) \leq \min_{F,V}\sum_{a} F(a)\,\mathfrak{C}\left[\sum_{\mathbf{x} \in \mathcal{X}^{L+1}} V(x_1|a)\cdots V(x_{L+1}|a) \mu_{\mathbf{x}}(\widetilde{\bm{\alpha}})\right]\,,
\end{equation}
where the minimum is over all types and conditional types $F$ and $V$ such that $FV=Q$ and $I(F,V) \leq R$. Even if, at first glance, the use of the upper concave envelope in the bound might look like a naive way of eluding a technical difficulty, it must instead be noted that the bound is tight for constant composition codes for\footnote{Here we mean codes with a subexponential number of codewords.} $R=0$. In fact, when $R=0^+$, condition $I(F,V)=0$ implies that $V(x|a) = Q(x)$ for every $a$. Therefore, \eqref{ELBoundQ} becomes
\begin{equation}
E_L(0^+,Q) \leq \mathfrak{C}\left[\sum_{\mathbf{x} \in \mathcal{X}^{L+1}} Q(x_1)\cdots Q(x_{L+1}) \mu_{\mathbf{x}}(\widetilde{\bm{\alpha}})\right]\,,
\end{equation}
and the right hand side turns out to be the true value of $E_L(0,Q)$ (see \cite[Prob. 10.22]{csiszar1}).

To conclude, using the well-known fact that from any code one can extract a constant composition subcode with rate asymptotically equal to that of the original code, one can maximize the bound \eqref{ELBoundQ} over all compositions $Q$, to obtain a bound on the reliability function for any code of rate $R$:
\begin{equation}
E_L(R) \leq \max_Q \min_{F,V}\sum_{a} F(a)\,\mathfrak{C}\left[\sum_{\mathbf{x} \in \mathcal{X}^{L+1}} V(x_1|a)\cdots V(x_{L+1}|a) \mu_{\mathbf{x}}(\widetilde{\bm{\alpha}})\right]\,,
\end{equation}
where the maximum is over all probability distributions on $\mathcal{X}$, and the minimum is over all types and conditional types $F$ and $V$ such that $FV=Q$ and $I(F,V) \leq R$. Once again, when $R\to0$ this bound becomes equal to \eqref{FinalBound}.

\section*{Acknowledgments}
This research was supported by the Italian Ministry of Education under grant PRIN 2015 D72F16000790001. We thank Yury Polyanskiy for pointing out reference \cite{blinovsky3} and the use of \cite{komlos1} therein.

\ifCLASSOPTIONcaptionsoff
  \newpage
\fi

\end{document}